\documentclass[runningheads]{llncs}

\usepackage{graphicx}
\usepackage{subcaption}
\usepackage{booktabs}
\usepackage{color,xcolor}
\usepackage{fixltx2e}
\usepackage{booktabs}
\usepackage[T1]{fontenc}
\usepackage{setspace}

\graphicspath{{figs/}}

\begin{document}

\title{Improving Software Requirements Prioritization through the Lens of Constraints Solving}
\titlerunning{Improving Software Requirements Prioritization}
%
\author{Jonathan Winton\inst{1} \and
Francis Palma\inst{2}\orcidID{0000-0001-7092-2244}}
\authorrunning{Jonathan Winton et al.}
%
\institute{Department of Computer Science, Linnaeus University, Kalmar, Sweden\\
\email{jw22gh@student.lnu.se}\\
\and
Faculty of Computer Science, University of New Brunswick, NB, Canada\\
\email{francis.palma@unb.ca}}
\maketitle              

\begin{abstract}
Requirements prioritization is a critical activity during the early software development process, which produces a set of key requirements to implement. The prioritization process offers a parity among the requirements based on multiple characteristics, including end-users' preferences, cost to implement, and technical dependencies. This paper presents an interactive method to requirements prioritization that leverages the pairwise comparisons and a constraint solver. Our method employs an interactive accumulation of knowledge from the requirements analyst when the relative priority among the requirements cannot be determined based on the existing knowledge from the requirements documents. The final ranking of the requirements is produced via the constraint solver and interactive pairwise comparisons. We evaluate the proposed method using the requirements from a real healthcare project. The proposed prioritization method relying on a constraint solver outperforms state-of-the-art interactive prioritization methods in terms of effectiveness and robustness to analyst's errors.
\keywords{Constraints Satisfaction \and Requirements \and Prioritization \and Interactive.}
\end{abstract}

\section{Introduction}\label{sec:Introduction}
Requirements prioritization determines the best set of requirements to implement and deliver in a certain release. Software projects usually have more candidate requirements to realize within several constraints like time and cost. The aim of the requirements prioritization process is to identify the most relevant requirements by determining the critical ones from the other trivial requirements \cite{aurum2003fundamental}. The support offered by the requirements prioritization process may include (i) determine the core requirements; (ii) decide an ordered, optimal set of requirements to implement and deliver; (iii) decide a subset of the requirements to develop a minimum viable product; (iv) manage the conflicting requirements in terms of technical dependency \cite{sommerville1997re,wiegers2013software}.

In the literature, a number of methods have been introduced to prioritize requirements \cite{in2002multi,karlsson1996software,karlsson1997cost,khan2016repizer,lauesen2002software,leffingwell2000managing,palma2011using,perini2012machine,tonella2013interactive}. Some methods, e.g., \cite{leffingwell2000managing} use the models based on the weighted requirements properties including cost, value, and effort. These methods are generally based on \textit{a priori} knowledge obtained prior to any development experience of the requirements (\textit{ex-ante}). In contrast, for the \textit{ex-post}, ordering of requirements is produced based on the characteristics of the specific requirements set without any predefined models (\textit{a posteriori}) to deduce requirements ordering. Methods that employ \textit{a posteriori} knowledge to prioritize requirements include \cite{in2002multi,karlsson1996software,karlsson1997cost,khan2016repizer,lauesen2002software,palma2011using,perini2012machine,tonella2013interactive}.

The most common methods for requirements prioritization include bubble sort \cite{karlsson1998evaluation}, cumulative voting or 100-dollar test \cite{leffingwell2000managing}, and top-ten requirements \cite{lauesen2002software}. In the cumulative voting method, stakeholders involved in prioritization are given a value of imaginary units (e.g., 100 dollars or points). They are asked to assign certain units to the requirements where the total value for a requirement indicates its priority \cite{leffingwell2000managing}. In the bubble sort method, requirements are compared pairwise manually, and this becomes infeasible for a higher number of requirements in a project. Unlike the AHP (Analytic hierarchy process) \cite{rodriguez2002models}, the bubble sort method does not concern the extent to which one requirement is more important. In a simple top-ten requirements method, stakeholders determine their top ten requirements without considering any internal order, and then the list is consolidated. The primary concerns for these methods are the scalability and applicability in a large, complex system with many requirements.

From the prioritization methods in the \textit{ex-post} category, the AHP \cite{rodriguez2002models} is widely recognized that exploits an exhaustive pairwise comparison to obtain requirements analyst's knowledge associated with the ordering of the requirements. However, exploiting all possible requirements pairs and evaluating them introduce the scalability problem with the AHP-based methods. Few methods are proposed, e.g., \cite{avesani2005facing,harker1987incomplete} to address the AHP's scalability problem. The key to handling the scalability problem is to reduce the number of pairwise comparisons while ordering performance is also considered. 

To handle scalability problems, methods like CBRank \cite{avesani2005facing} can be useful that leverage machine learning and can encode simple constraints like user priority for the requirements. Although CBRank is interactive, it does not support encoding additional constraints like dependencies among the requirements or the elicited analyst knowledge. This problem was resolved by other interactive approaches like Incomplete AHP (IAHP) \cite{harker1987incomplete}, Interactive Genetic Algorithm (IGA) \cite{tonella2013interactive}, and Yices SMT solver-based method \cite{palma2011using}. These methods could also handle the scalability problem well, i.e., with fewer analyst input, they still outperform state-of-the-art non-interactive methods. However, the rankings of the requirements produced by the state-of-the-art interactive methods are not the best. Therefore, in this paper, we aimed to improve the prioritization further.


This paper proposes an \textit{ex-post} method to synthesize requirements ordering considering various constraints. The proposed method employs pairwise comparison in the form of preference from the requirements analysts, referred to as \textit{user interaction}. During the user interaction session, the primary purpose is to extract project and requirements-related knowledge from the analysts, given their experience and skills in requirements engineering. Interaction sessions are useful when the existing knowledge of requirements is not adequate to decide on the relative priority for a set of requirements pairs. Our proposed method relies on a well-known constraint solver developed by Microsoft (i.e., Z3 \cite{de2008z3}). The final goal is to reduce the effort by the analyst in terms of the number of pairwise elicitation while improving the prioritization ranking. The requirements pairs elicited by the analyst and the initial requirements knowledge (a.k.a., domain knowledge) combine as the \textit{constraints}. The elicitation and optimization are carried out simultaneously and can influence one another. The elicited constraints are constituted iteratively and incrementally.

In this paper, we answer three research questions:

\begin{itemize}
    \item RQ1 (\textit{Role of Interactio}n): Does the interactive Z3-based method improve the requirements prioritization compared to the non-interactive Z3-based method?
    \item RQ2 (\textit{Comparison}): Does the Z3-based method improve the requirements prioritization compared to the SMT-based, IAHP, and IGA methods?
    \item RQ3 (\textit{Robustness}): Is the Z3-based method more robust than the SMT-based, IAHP, and IGA methods with respect to the errors committed by the analyst during the interactive session?
\end{itemize}

Our proposed method further improves the prioritization than the state-of-the-art methods (IAHP \cite{harker1987incomplete}, IGA \cite{tonella2013interactive}, and Yices\footnote{https://yices.csl.sri.com}-based SMT solver \cite{palma2011using}) in terms of ranking. Results show that our prioritization algorithm based on a constraint solver, Z3, substantially outperforms previous methods (e.g., IAHP, IGA, and Yices-based SMT solver) with a similar effort in terms of the total number of elicited pairs by the analyst. Moreover, the ranking is improved using the constraint solver-based interactive method compared to the non-interactive version of the solution. We assessed the robustness of our Z3-based algorithm concerning analyst's decision-making errors.

This paper is structured as follows: Section \ref{sec:Background} provides the background for this work, Section \ref{sec:Method} presents the proposed method and our prioritization algorithm. Section \ref{sec:Experiments} details experimental evaluations of the proposed method on its effectiveness and robustness. We rely on a healthcare project as our case study. Finally, Section \ref{sec:Conclusions} concludes and outlines future work.

\section{Background}\label{sec:Background}
Here we introduce three state-of-the-art interactive requirements prioritization methods.

\subsection{Incomplete AHP (IAHP)} The AHP (Analytic Hierarchy Process) \cite{rodriguez2002models} method requires pairwise comparisons of the requirements. The number of comparisons that the user make is quadratic to the number of requirements, i.e., \texttt{N*(N-1)/2} with \texttt{N} requirements. In AHP, the analyst specifies an integer value to quantify the relative importance between two requirements. For example, if requirement \texttt{R1} is strongly important (e.g., an integer value is 5) than \texttt{R2}; then, the preference value between \texttt{R1} and \texttt{R2} is 5, and the preference value between \texttt{R2} and \texttt{R1} is \texttt{1/5} (or \texttt{0.2}). The preference value for the less preferred requirement is reciprocal. The AHP builds a comparison matrix. Once all the comparisons are made, the ranking of the requirements can be synthesized from the principal eigenvector of the matrix. The elements of the calculated matrix decide the final requirements ordering. However, AHP suffers from the scalability problem. 
As a variant of AHP, the IAHP (Incomplete AHP) \cite{harker1987incomplete} operates with incomplete information in the form of pairwise comparisons from the analyst. Thus, IAHP overcomes the scalability problem by reducing the number of comparisons while balancing the ranking performance and the analyst's effort. This can be achieved by predicting the next most promising pair to be elicited with a good approximation towards the final ranking.

\subsection{Interactive Genetic Algorithm (IGA)} An IGA, applying a genetic algorithm, seeks to reduce the disagreement between a total order of ranked requirements and the domain knowledge in the form of various constraints. In \cite{tonella2013interactive}, constraints either come with the requirements or from the experience and skills of the analyst. New constraints (i.e., analyst knowledge) are introduced to the prioritization process when the existing domain knowledge is not enough to decide the relative importance of the requirements. 
For example, if two requirements orderings (i.e., chromosomes) \texttt{<R1, R5, R4, R3, R2>} and \texttt{<R1, R4, R5, R2, R3>} cannot be distinguished due to their equal fitness against the existing domain knowledge, input from the analyst is introduced. The analyst is involved in eliciting the pairs \texttt{(R4, R5)} and \texttt{(R2, R3)} because these contradict the example orderings. Involving analyst knowledge will lead to introducing new constraints helping to further discriminate the orderings. The operators related to the genetic algorithm (i.e., selection, crossover, and mutation) and user interaction help generate new discriminative orderings with lower disagreement. The prioritization process terminates upon reaching a low threshold disagreement or a set timeout, or exceeding the elicitation budget.

\subsection{SMT Solver-based Method} In an SMT solver-based (e.g., Yices) method \cite{palma2011using}, the requirements prioritization is encoded as a \texttt{MAX-SAT} problem. The encoding of the constraints is represented as inequality relations. For example, the requirement \texttt{R4} has a higher user priority than \texttt{R1} or the requirement \texttt{R1} has a technical dependency on \texttt{R4}. In that case, it can be encoded as \texttt{R1 < R4} and can be expressed in an SMT solver as an assertion \texttt{(assert+ (< (R1) (R4)) 1)}, with a default weight of \texttt{1}. The solver can have numerous such inequality relations between the requirements. The solver then finds the minimum number of violated relations (i.e., the cost) and an ordering of the requirements. If there are multiple ordering of the requirements with the same cost, an analyst is involved in eliciting pairs to discriminate the orderings further. The analyst's input creates new domain knowledge, passed to the solver incrementally and iteratively. The prioritization process terminates when the solver returns a unique solution with a minimum cost or the elicitation quota exceeds.

\section{Our Proposed Method}\label{sec:Method}

We rely on a constraint solver to find an ordered set of requirements. Using our prioritization method, we aim to minimize disagreement between final ranked requirements and existing domain knowledge. Domain knowledge is extracted from the requirements documents or based on the analyst's input.

\begin{figure}[t!]
\begin{subfigure}[t]{0.25\textwidth}
         \centering
         \includegraphics[width=1\textwidth]{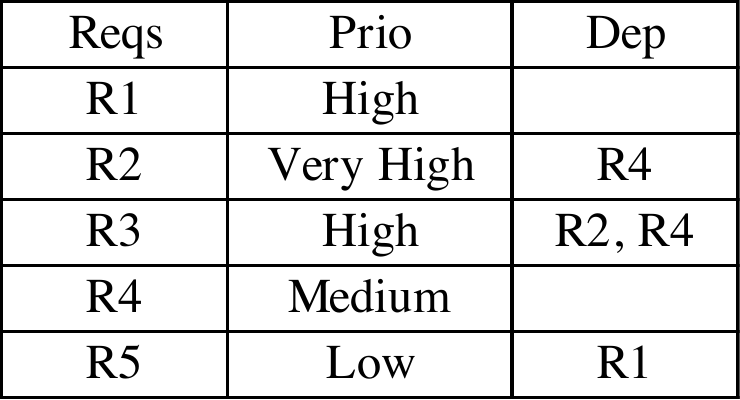}
         \caption{Requirements}\label{Requirements}
     \end{subfigure}
     \hfill
     \begin{subfigure}[t]{0.25\textwidth}
         \centering
         \includegraphics[width=0.5\textwidth]{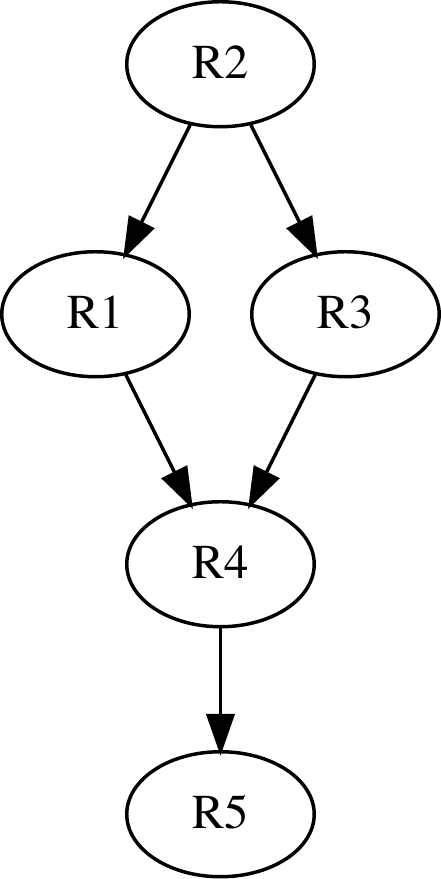}
         \caption{Prio}\label{Prio}
     \end{subfigure}
    \hfill
     \begin{subfigure}[t]{0.25\textwidth}
         \centering
         \includegraphics[width=0.6\textwidth]{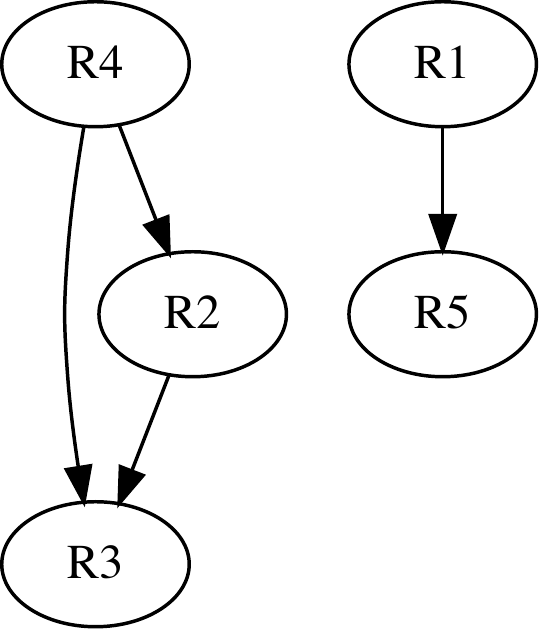}
         \caption{Dep}\label{Dep}
     \end{subfigure}
        \caption{Example requirements with priority and dependencies.}
        \label{fig:requirements}
\end{figure}

Figure \ref{fig:requirements} shows five requirements with the end-user's priority and technical dependencies among them. The constraints on the requirements can be expressed via directed graphs. In such a graph, the dependency between two requirements can be represented as an edge. For example, in Figure \ref{Dep}, we have an edge from \texttt{R1} to \texttt{R5}, i.e., there is a dependency between \texttt{R1} and \texttt{R5} as Figure \ref{Requirements} shows that \texttt{R5} depends on \texttt{R1}. Thus, the developer should implement \texttt{R1} before he decides to implement \texttt{R5}. Indeed, there can be multiple such dependencies that constitute the \textit{Dep} graph as in Figure \ref{Dep}. In addition, Figure \ref{Prio} shows that we have an edge from \texttt{R2} to \texttt{R1}, i.e., \texttt{R2} should be implemented before \texttt{R1} because \texttt{R2} has a higher end-user priority than \texttt{R1}, as shown in Figure \ref{Requirements}. In the \textit{Prio} graph, we have requirements in several layers, each layer representing a priority level, as shown in Figure \ref{Prio}. In this form, constraints can be encoded to directed and acyclic graphs. Edges can be weighted to indicate that certain relations or constraints are relatively more important. This paper uses the default weight of 1 for the edges. The constraint solver may ignore default weighted constraints (a.k.a., retractable constraints), searching for the solution with minimum cost). An infinite weight can be used for relations or constraints that must hold. However, the solver cannot ignore constraints with infinite weights, a.k.a., non-retractable constraints.

\subsection{The \texttt{CHECK-SAT} Problem}

With the constraint graphs, we transform the prioritization problem to a \texttt{CHECK-SAT} problem \cite{de2008z3}. A constraint solver decides on placing requirements in order, considering that each requirement is assigned a unique position (i.e., no duplicate requirements in a solution) to maximize the weights from the satisfied and retractable constraints with a minimum cost of unsatisfied constraints. The constraints from the directed graphs are encoded as an inequality relation, e.g., \texttt{(assert-soft (R4 < R1) :weight 1)} for one of the first two edges from the \textit{Prio} graph (see Figure \ref{fig:requirements}). The solutions returned on the \texttt{CHECK-SAT} problem are a set of requirements orderings violating the minimum number of retractable constraints (i.e., minimum cost). Figure \ref{fig:encoding} shows the encoded prioritization problem using the Z3 constraint solver.

\begin{figure}[t!]
\begin{spacing}{0.85}
\begin{small}
\begin{tabular}{p{11.5cm}}
\bottomrule
\texttt{(declare-datatypes () ((R (mk\_R (key Int) (var1 Int) (var2 Int)))))}\\
\texttt{(declare-const R\_instances (Array Int R))}\\
\texttt{(declare-fun j () Int)}\\
\texttt{(assert (forall ((i Int))  (implies (distinct i j)}\\
\hspace{8mm}\texttt{(distinct (key (select R\_instances i))}\\
\hspace{16mm}\texttt{(key (select R\_instances j))))))}\\
\texttt{;Prio}\\
\texttt{(assert-soft (R4 < R1) :weight 1)}\\
\texttt{(assert-soft (R5 < R1) :weight 1)}\\
\texttt{(assert-soft (R1 < R2) :weight 1)}\\
\texttt{(assert-soft (R3 < R2) :weight 1)}\\
\texttt{(assert-soft (R4 < R2) :weight 1)}\\
\texttt{(assert-soft (R5 < R2) :weight 1)}\\
\texttt{(assert-soft (R4 < R3) :weight 1)}\\
\texttt{(assert-soft (R5 < R3) :weight 1)}\\
\texttt{(assert-soft (R5 < R4) :weight 1)}\\
\texttt{;Dep}\\
\texttt{(assert-soft (R2 < R4) :weight 1)}\\
\texttt{(assert-soft (R3 < R2) :weight 1)}\\
\texttt{(assert-soft (R3 < R4) :weight 1)}\\
\texttt{(assert-soft (R5 < R1) :weight 1)}\\
\\
\texttt{(check-sat)}\\
\bottomrule
\end{tabular}
\end{small}
\end{spacing}
    \caption{Encoding of the constraints in Figure \ref{fig:requirements} for the Z3 Constraint Solver.}
    \label{fig:encoding}
\end{figure}

\begin{table}[t!]
\begin{spacing}{0.85}
    \centering
    \begin{small}
    \begin{tabular}{c c c}
    \toprule
     Solution ID  & Requirements Order & Disagreement \\
    \midrule
     \texttt{Pr1} & \texttt{$<$R2, R1, R4, R3, R5$>$} & 2 \\
     \texttt{Pr2} & \texttt{$<$R2, R3, R1, R4, R5$>$} & 2 \\
     \texttt{Pr3} & \texttt{$<$R2, R1, R3, R4, R5$>$} & 2 \\
    \bottomrule
    \end{tabular}
    \end{small}
    \caption{Requirements ranking with minimum disagreements.}
    \label{tab:exampleresult}
    \end{spacing}
\end{table}

We obtain multiple solutions by running the Z3 constraint solver in the form of ranked requirements with minimum cost, as shown in Table \ref{tab:exampleresult}. Prior to that, we encode the graphs shown in Figure \ref{fig:requirements} as retractable assertions, i.e., \texttt{assert-soft} in Figure \ref{fig:encoding}. The disagreement of 2 in Table \ref{tab:exampleresult}, the cost returned by the constraint solver, is the number of constraints not recognized by the positions of the requirements in the ranked order decided by the constraint solver. The last two solutions, \texttt{Pr2} and \texttt{Pr3}, do not disagree with the \textit{Prio} graph. In contrast, they disagree with the \textit{Dep} graph for two edges, \texttt{R3$\rightarrow$R4} and \texttt{R2$\rightarrow$R4}. However, the first solution \texttt{Pr1}, violates \texttt{R2$\rightarrow$R4} and \texttt{R4$\rightarrow$R3} against \textit{Dep} and \textit{Prio} graphs.

\begin{table}[t!]
\begin{spacing}{0.85}
    \centering
    \begin{small}
    \begin{tabular}{c c l}
    \toprule
    Solution Pairs && Set of Pairs in Disagreement \\
    \midrule
     \texttt{Pr1, Pr2} & & \texttt{<R1, R3>, <R3, R4>} \\
     \texttt{Pr1, Pr3} & & \texttt{<R3, R4>} \\
     \texttt{Pr2, Pr3} & & \texttt{<R1, R3>} \\
    \bottomrule
    \end{tabular}
    \end{small}
    \caption{Pairs in Disagreement among the obtained Solutions.}
    \label{tab:pairsinDisagree}
    \end{spacing}
\end{table}

Since we have multiple solutions with the same cost, we can leverage the benefit of analyst's knowledge and experience, which could further discriminate the solutions. From the three solutions we obtained, the requirements pairs: \texttt{<R1, R3>} and \texttt{<R3, R4>} are in disagreement between solutions \texttt{Pr1} and \texttt{Pr2}. The pair \texttt{<R3, R4>} is in disagreement between solutions \texttt{Pr1} and \texttt{Pr2}, while between the solutions \texttt{Pr2} and \texttt{Pr3}, they disagree for the requirements pair \texttt{<R1, R3>}.

At this stage, the analyst's role is critical to decide on the relative importance of the requirements pairs that are in disagreement. The analyst's feedback is translated into another graph \textit{Eli} in the form of constraints. The \textit{Eli} graph is initially empty, and an updated \textit{Eli} graph will be part of the \texttt{CHECK-SAT} problem during the next iteration. For the above pairs, for example, the analyst may decide for the two unique pairs in Table \ref{tab:pairsinDisagree} that \texttt{R4} is more important than \texttt{R3} (for the pair \texttt{<R3, R4>}) and \texttt{R3} is more important than \texttt{R1} (for the pair \texttt{<R1, R3>}). Thus, the \textit{Eli} graph is updated with two more edges, \texttt{R4$\rightarrow$R3} and \texttt{R3$\rightarrow$R1}, and so the encoded constraints for the \texttt{CHECK-SAT} problem in the next iteration. The analyst may be undecided, for which no new edge is introduced in \textit{Eli} graph.

With the new constraints graph \textit{Eli}, the previously existed domain knowledge \textit{Prio} and \textit{Dep}, and the non-retractable assertions for requirements positioning required by the constraint solver, the solutions are recomputed using \texttt{CHECK-SAT}. The new solutions would be more discriminative than the previous ones, thanks to the analyst's role with his knowledge and experience in discriminating the tied solutions in terms of disagreement.

\subsection{Z3 Constraint Solver-based Prioritization Algorithm}\label{subsec:Algorithm}

Here we outline the requirements prioritization algorithm that realizes the method presented in Section \ref{sec:Method}. We also introduce the term \textit{disagreement} and formalize the problem. The disagreement is calculated between two orders of requirements where both are partial orders, or at least one is a total order. The disagreement between two (partial or total) requirements orders \textit{Ord$_{1}$} and \textit{Ord$_{2}$}, defined on a set of requirements \texttt{R} is the pairs of requirements in transitive closure in \textit{Ord$_{1}$} that are opposite in \textit{Ord$_{2}$}. Consequently, the disagreement value is the size of the set \textit{disagreement(Ord$_{1}$,Ord$_{2}$)} (see Equation \ref{eq1}) \cite{5635176}.

\begin{equation}\label{eq1}
disagreement(Ord_{1},Ord_{2})= \{(R_{i},R_{j}) \in Ord_{1} | (R_{j},R_{i}) \in Ord_{2}\}
\end{equation}

In this paper, the requirements prioritization problem is formulated as the problem of specifying integer values for the requirements positions \texttt{Pos={1,...,N}}, with \texttt{N} requirements, to an integer array \texttt{A} in such a way that all requirements hold a unique index, i.e., no two or more requirements have same positions in the array. We can formalize this as follows:

\begin{equation}\label{eq2}
\forall \hspace{1mm} i \in Pos, j \in Pos : A_{i} \in Pos, A_{j} \in Pos, i \ne j \Rightarrow A_{i} \ne A_{j}
\end{equation}

Considering the placement constraints as in Equation \ref{eq2}, solutions with the minimum number of unsatisfied constraints are desired. This can be obtained by encoding the constraints represented as graphs in the form of retractable inequality constraints, i.e., \texttt{R1 < R2} if an edge exists from \texttt{R2} to \texttt{R1} (see Figure \ref{fig:requirements}). The disagreement refers to the least number of inequality relations retracted in the \texttt{CHECK-SAT} solution by the constraint solver. We iterate the solver to enumerate all possible solutions at minimum disagreement or cost.

\begin{table}[t!]
\begin{spacing}{1}
\begin{small}
\begin{tabular}{p{11.7cm}}
\toprule
\noindent \textbf{Algorithm 1:} Requirements Prioritization Algorithm using a Constraint Solver.\\
\toprule
\noindent \texttt{\textbf{IN:} R: a set of requirements} \\

\noindent \texttt{\textbf{IN:} ord$_1$,..,ord$_k$: partial orders (constraints) on R (ord$\subseteq$R$\times$R)}\\

\noindent \texttt{\textbf{OUT:} (R$_1$,..,R$_n$): ordered list of requirements} \\

\noindent 1: \texttt{\textit{Solutions} := $\emptyset$} \\

\noindent 2: \texttt{\textit{eliPair} := 0} \\

\noindent 3: \texttt{\textit{maxEliPair} := MAX\_ELI\_Pair (default 100)} \\

\noindent 4: \texttt{\textit{eliOrd} := $\emptyset$} \\

\noindent 5: \texttt{\textbf{DO}} \\

\noindent 6: \hspace{4mm}\texttt{Solutions := CHECK-SAT(ord$_{1}$,..,ord$_{k}$, \textit{eliOrd})} \\

\noindent 7: \hspace{4mm}\texttt{\textbf{IF} $\mid$Solutions$\mid$ $\neq$ 1 \textbf{THEN}} \\

\noindent 8: \hspace{8mm}\texttt{\textit{eliOrd} := \textit{eliOrd} $\cup$ elicit pairs up to \textit{maxEliPair}} \\

\noindent 9: \hspace{8mm}\texttt{increase \textit{eliPair} by the number of elicited pairs} \\

\noindent 10: \texttt{\textbf{WHILE} $\mid$Solutions$\mid$ $\neq$ 1 \textbf{AND} \textit{eliPair} $\le$ \textit{maxEliPair}} \\

\noindent 11: \texttt{\textbf{RETURN} Pr$_{min}$, a requirements list from \textit{Solutions} with min} \texttt{disagreement.} \\
\bottomrule
\end{tabular}
\end{small}
\end{spacing}
\end{table}

Algorithm 1 presents the pseudocode of our proposed requirements prioritization algorithm. Our algorithm has two sets of initial inputs: (1) a set of requirements and (2) a set of one or more partial requirements order, e.g., \textit{Prio} and \textit{Dep} in Figure \ref{fig:requirements}. The partial requirements order can be obtained from the requirements documents and various stakeholders. Initially, we have an empty set of solutions (\texttt{\emph{Solutions}}, Step 1), and no pairs have been elicited, i.e., \texttt{\emph{eliPair}} is zero (Step 2). In Step 3, we set \texttt{\emph{maxEliPair}} as the upper limit for the pairwise comparisons made by the requirements analyst. The default value for \texttt{\emph{maxEliPair}} is 100, other values are common (e.g., 25 and 50). The initial graph for the elicited pairs, \texttt{\emph{eliOrd}}, is also empty, as in Step 4.

Inside the iteration, in Step 6, the designated constraint solver is invoked. The constraint solver is fed with the retractable assertions retrieved from the constraints graphs. These graphs are turned into inequality assertions, as discussed above. Moreover, the constraint solver includes the non-retractable assertions all the time as part of the \texttt{CHECK-SAT} problem, as shown in Equation \ref{eq2}. In Step 7, if the constraint solver returns multiple solutions with minimum cost, the requirements analyst gets involved and makes pairwise comparisons (Step 8), for which the \texttt{\emph{eliPair}} is updated by the number of elicited pairs (Step 9). The analyst's input is encoded as \texttt{\emph{eliOrd}} in the form of a constraint graph. This new knowledge is used as retractable assertions during the next iteration by the constraint solver in the algorithm. It is important to note that the pairwise comparisons continue until the \texttt{\emph{eliPair}} is less than the \texttt{\emph{maxEliPair}} (Step 8). 

As part of Step 6 in our algorithm, the constraint solver lists all possible solutions with the least possible cost, internally, which is done by introducing an assertion to neutralize the previous solutions, and, thus, searches for new solutions to the reformulated problem. Thus, each new solution is created from the retractable assertions from the constraints and non-retractable assertions, as in Equation \ref{eq2}, with the negation of the previous solution. In Step 6, the constraint solver continues until it exhausts all possible solutions with the minimum cost. The prioritization algorithm loops until it reaches the maximum number of permitted elicited pairs \texttt{\emph{maxEliPair}} or until the constraint solver returns a unique solution with the minimum cost, i.e.,  disagreements against the constraints graphs (Step 10).

The uniqueness of our proposed algorithm comes from the input from the analyst, thanks to his knowledge and experience on the project. When the constraint solver reaches a plateau where multiple candidate solutions have the minimum cost with no further discriminatory knowledge, the analyst plays a critical role in driving the requirements prioritization process, making it interactive.

\section{Experiments}\label{sec:Experiments}

We perform a suite of experiments using the Z3 constraint solver-based prioritization algorithm using the requirements from the ACube (Ambient Aware Assistance) project \cite{andrich2010acube}. ACube is a system developed for an elderly care facility to assist the staff. The project has 49 technical requirements and four macro scenarios. The scenarios include (1) FALL: locate and track residents to identify falls (26 requirements), (2) ESCAPE: locate and track residents to detect an escape from the facility (23 requirements), (3) MONITOR: identify dangerous behaviors by the residents (21 requirements), (4) ALL: a combination of three macro scenarios (49 requirements). Two more constraints are also gathered during requirements elicitation: priority (\textit{Prio}) and dependency (\textit{Dep}). We assess the performance of our method in terms of two metrics: \textit{disagreement} and \textit{average distance}. Disagreement is calculated as the number of pairs that appear reverse in the gold standard (GS). The average distance is the average index displacement for each requirement against the GS. We leverage the availability of the GS defined by the architect of the ACube project.


\subsection{Results}

\begin{table}[t!]
    \centering
    \setlength{\tabcolsep}{3.5pt}
    \begin{scriptsize}
    \begin{tabular}{cccccccccccccccc}
    \toprule
Elicited & Actual & \multicolumn{2}{c}{\textbf{Z3}}	 &	\multicolumn{2}{c}{\textbf{Z3(5\%)}} & \multicolumn{2}{c}{\textbf{Z3(10\%)}}	 &	\multicolumn{2}{c}{\textbf{Z3(20\%)}}	 & \multicolumn{2}{c}{\textbf{SMT}} &	\multicolumn{2}{c}{\textbf{IGA}}	 &	\multicolumn{2}{c}{\textbf{IAHP}} \\	
 pairs&		 &	Dis	 &	AD	 &	Dis	 &	AD	 &	Dis	 &	AD	 &	Dis	 &	AD	 &	Dis	 &	AD	 &	Dis	 &	AD	 &	Dis	 &	AD \\
\midrule

25 &	25	 &	83	 &	2.9	 &	87	 &	2.98	 &	87	 &	3.06	 &	89	 &	3.06	 &	92	 &	3.18	 &	124	 &	4.06	 &	478	 &	13.6 \\ 

50 &	50	 &	78	 &	2.78	 &	82	 &	2.94	 &	85	 &	3.02	 &	88	 &	3.06	 &	90	 &	3.14	 &	120	 &	3.82	 &	208	 &	6.3 \\ 

100 &	84	 &	73	 &	2.78	 &	75	 &	2.86	 &	79	 &	2.94	 &	83	 &	3.02	 &	73	 &	2.78	 &	114	 &	3.69	 &	187	 &	5.75 \\ 

\bottomrule
\end{tabular}
\end{scriptsize}
    \caption{Disagreement (Dis) and Average Distance (AD) among Z3, SMT, IGA, and IAHP for various Elicited Pairs at various Error Rates for ALL Scenario.}
    \label{tab:Results}
\end{table}

\vspace{2mm}
\noindent \textbf{RQ1 (Role of Interaction)} Analyst's knowledge can play a critical role part in improving requirements prioritization. Our first research question assesses the effectiveness of analyst knowledge by comparing the requirements prioritization produced without the analyst's involvement and with increased participation from the analyst, i.e., a higher number of elicited pairs.

\begin{figure}[t!]
\begin{subfigure}[t]{0.48\textwidth}
         \centering
         \includegraphics[width=1\textwidth]{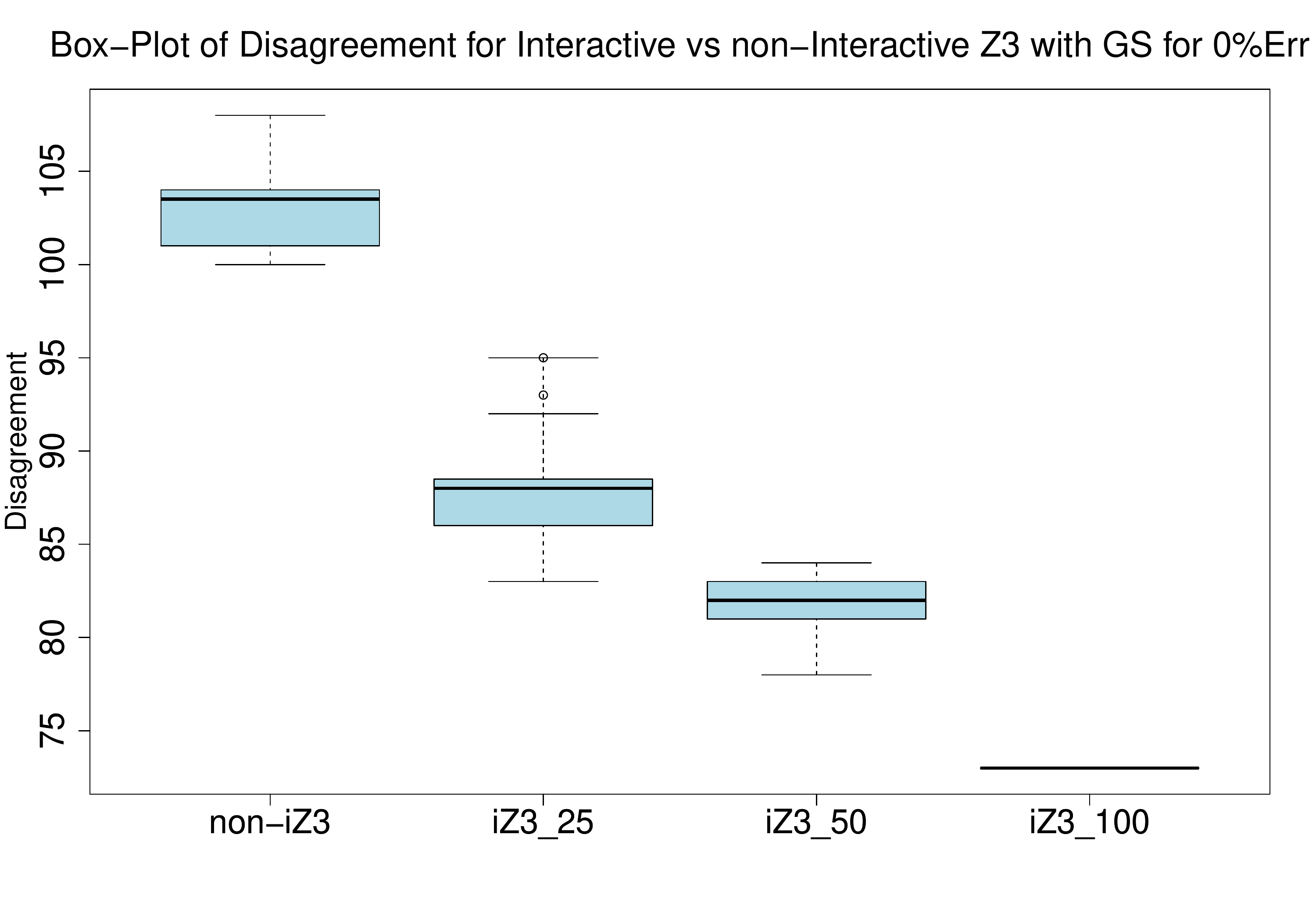}
     \end{subfigure}
     \hfill
     \begin{subfigure}[t]{0.51\textwidth}
         \centering
         \includegraphics[width=1\textwidth]{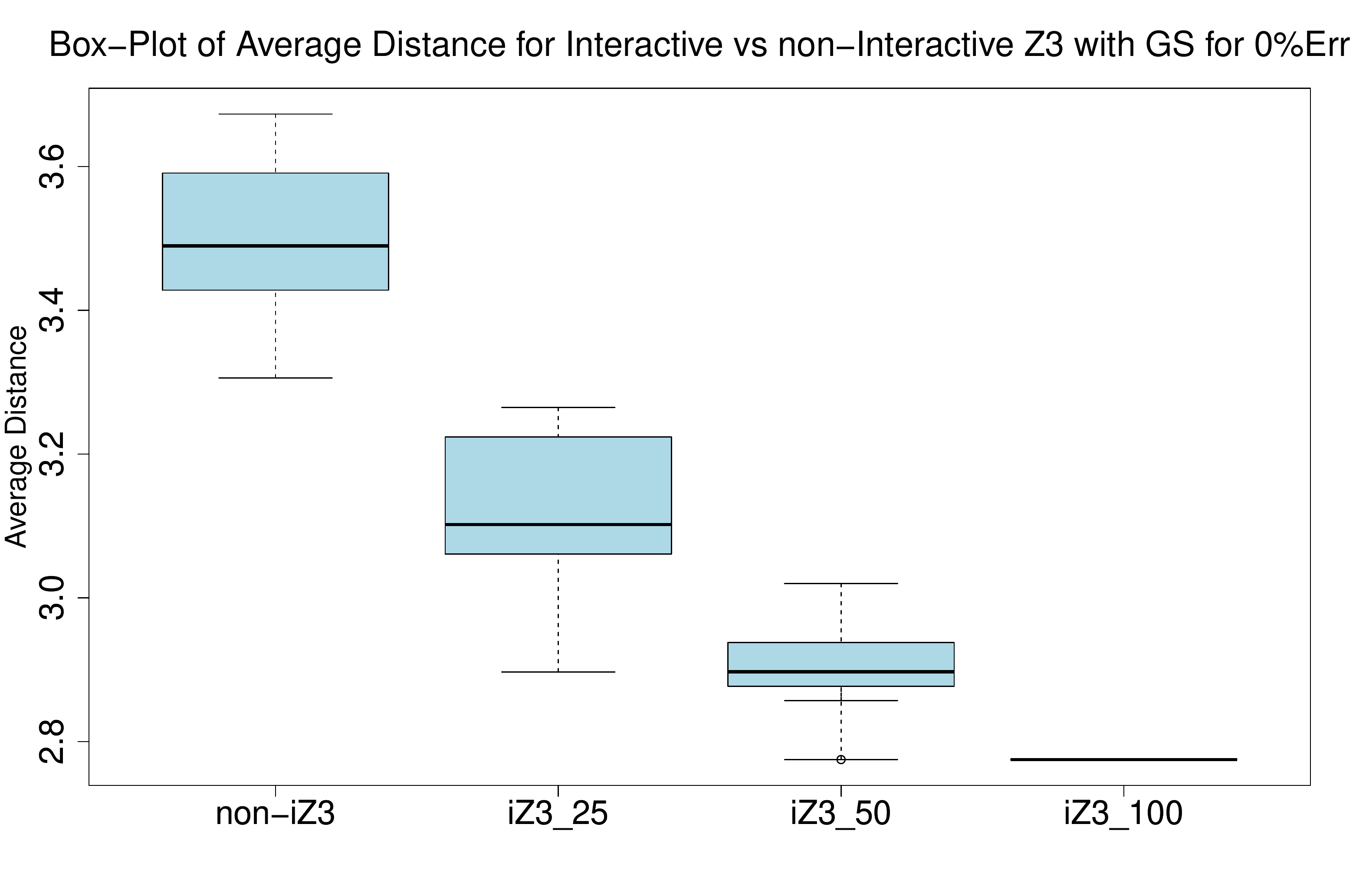}
     \end{subfigure}
    \hfill
        \caption{The Minimum Disagreement (left) and Average Distance (right) against GS for Interactive and Non-interactive Z3, different numbers of elicited pairs, ALL scenario.}
        \label{fig:RQ1}
\end{figure}

Figure \ref{fig:RQ1} depicts the performance of our Z3-based prioritization algorithm comparing the non-interactive and interactive methods for ALL scenario. In Figure \ref{fig:RQ1} (left), we show the disagreement, while Figure \ref{fig:RQ1} (right) shows the average distance against the GS. In RQ1, while assessing the role of interaction, we use the value for \emph{\texttt{maxEliPairs}} as 0, 25, 50, and 100. A zero value for \emph{\texttt{maxEliPairs}} refers to the non-interactive version of our method, i.e., no pairwise comparisons. In Figure \ref{fig:RQ1}, the disagreement (so is the average distance) gets lower as we increase the number of pairwise comparisons. The median disagreement using the non-interactive algorithm is 103.5 over a set of 20 executions of our algorithm. On the other hand, the median disagreements using the interactive algorithm are 88, 82, and 73 when we elicit 25, 50, and 100 pairs, respectively. Thus, we are improving on the requirements prioritization for increased interactions. Similarly, for the average distance in Figure \ref{fig:RQ1} (right), the median of the average distance using the non-interactive algorithm is 3.49. With the interactive version of the method, the median average distance are 3.1, 2.9, and 2.78 when we elicit 25, 50, and 100 pairs, respectively. By eliciting up to 100 elicited pairs (in fact, 84, see Table \ref{tab:Results}), we could improve the ranking of the requirements by more than 40\%. We performed ANOVA tests to observe the statistical significance of performance difference between non-interactive and interactive methods. Our ANOVA test in Table \ref{tab:RQ1:test} confirms the significance of the improvement with a \emph{p-value}$<$0.05.

\begin{table}[t!]
    \centering
    \begin{scriptsize}
    \begin{tabular}{c c c}
    \toprule
    Methods	& Measures	& \textit{p-value} \\
    \midrule
    Z3-0Eli, Z3-25Eli, Z3-50Eli, Z3-100Eli	& \textit{Disagreement}	& \textit{p<0.05} \\
    Z3-0Eli, Z3-25Eli, Z3-50Eli, Z3-100Eli	& \textit{Average Distance}	& \textit{p<0.05} \\
    \bottomrule
    \end{tabular}
    \end{scriptsize}
    \caption{ANOVA test comparing Z3-based Solutions with different Elicited Pairs with 0\% Analyst Error.}
    \label{tab:RQ1:test}
\end{table}

\vspace{2mm}
\noindent \textbf{RQ2 (Comparison)} We assess the effectiveness of the Z3-based method in comparison with other state-of-the-art interactive methods, i.e., SMT-based Yices, IGA, and IAHP. We conjecture that with a certain number of elicited pairs, our Z3-based method improves the disagreement against the GS.

\begin{figure}[t!]
\begin{subfigure}[t]{0.49\textwidth}
         \centering
         \includegraphics[width=1\textwidth]{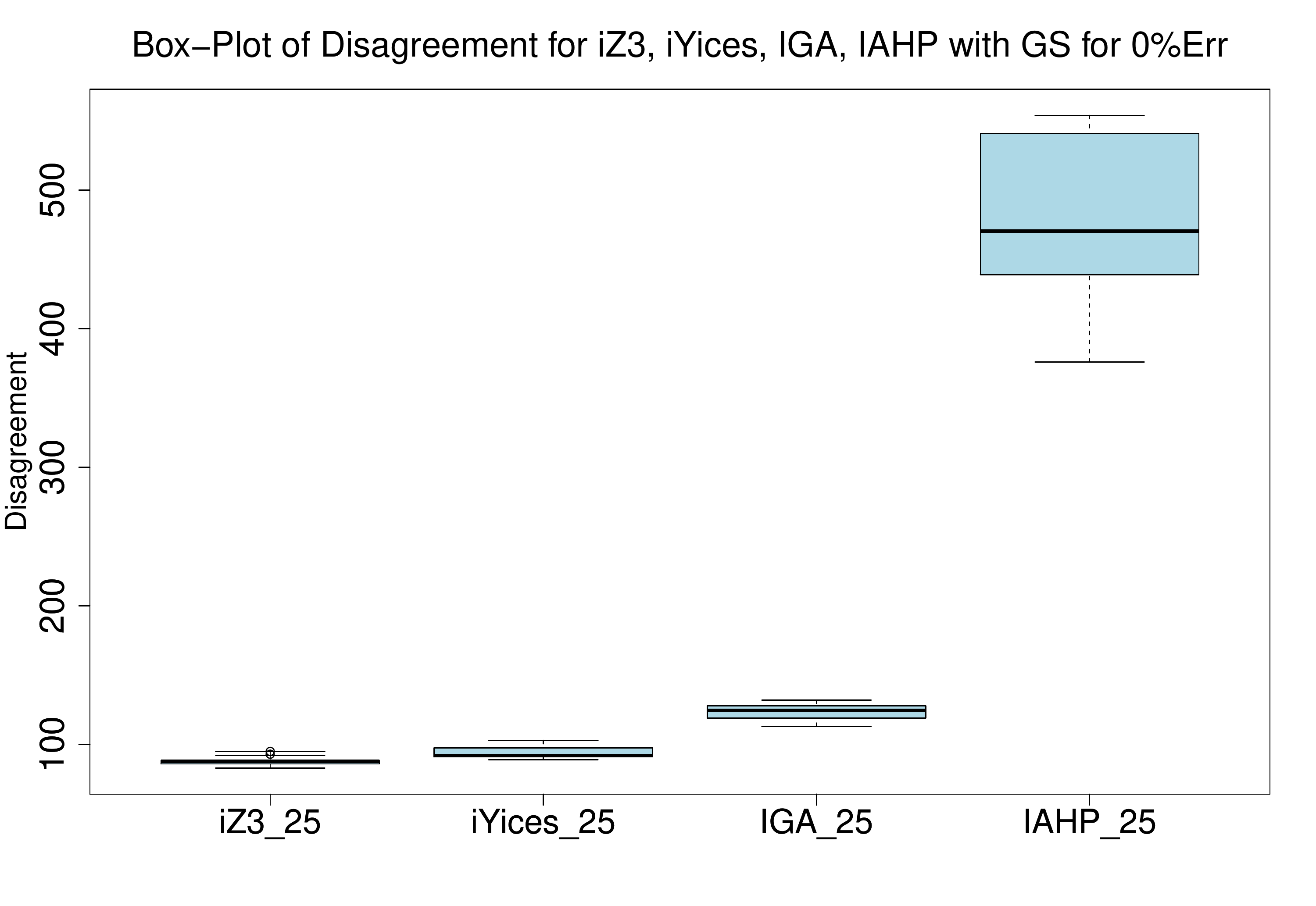}
     \end{subfigure}
     \hfill
     \begin{subfigure}[t]{0.5\textwidth}
         \centering
         \includegraphics[width=1\textwidth]{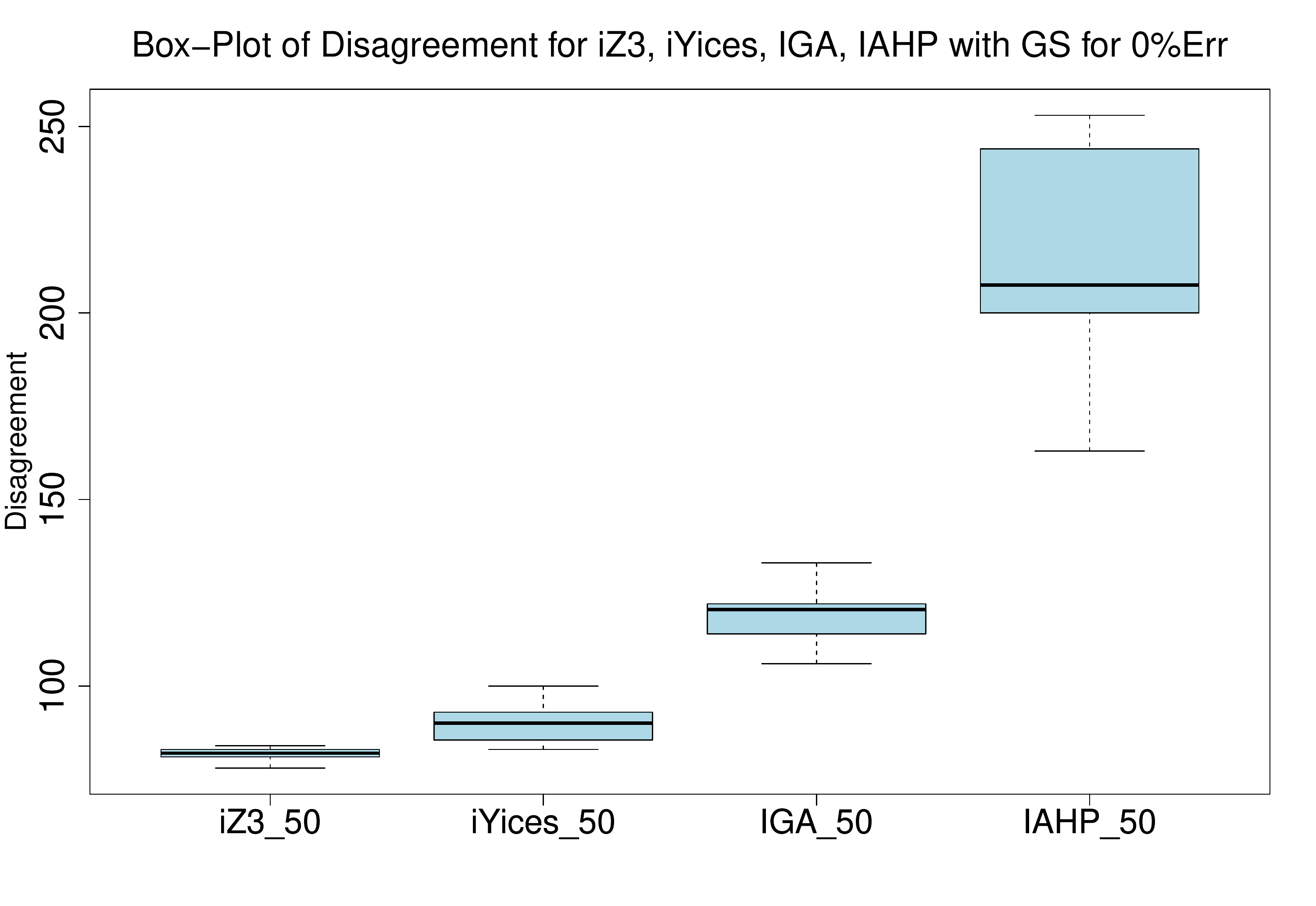}
     \end{subfigure}
        \caption{Disagreement against Gold Standard with 25, 50, and 100 pairs at most from the analyst for ALL Scenario.}
        \label{fig:RQ2}
\end{figure}

Figure \ref{fig:RQ2} shows the performance of our Z3-based method and compares it with other interactive methods (SMT, IGA, and IAHP) in terms of disagreement for 25, 50, and 100 pairwise elicitation for ALL scenario. Figure \ref{fig:RQ2} (left) shows that our Z3-based interactive method performs better than three other state-of-the-art interactive methods. Our Z3-based method performs close to the SMT-based method when we elicit 25 pairs. However, with more pairwise comparisons (i.e., 50), the Z3-based method clearly outperforms the SMT-based method and is profoundly better in disagreement than two other interactive methods (IGA and IAHP). More precisely, the median disagreement using Z3 and eliciting 25 pairs is 88, and with the same number of elicited pairs, SMT resulted in a disagreement of 92. Besides, the IGA and IAHP resulted with 124 and 478 as median disagreement, which is not comparable. Considering 50 elicited pairs, Z3 yields 82 as median disagreement, while SMT, IGA, and IAHP yielded 90, 120, and 208 for the same set of requirements and 50 elicited pairs. 

With a maximum of 100 elicited pairs, SMT and Z3 methods evaluate a plateau with a minimum disagreement of 73, as shown in Table \ref{tab:Results}. Also, the average distance for the requirements produced by SMT and Z3 are the same (i.e., 2.78) with max possible elicited pairs. Note that while we set the \emph{\texttt{maxEliPairs}} as 100, we could elicit 84 pairs, i.e., the constraint solver constrained the search space. We performed ANOVA tests to observe the statistical significance of differences and found that the disagreement values produced by state-of-the-art interactive methods are significantly higher than Z3 with a \emph{p-value}$<$0.05, as reported in Table \ref{tab:RQ2:test}.

\begin{table}[t!]
    \centering
    \begin{scriptsize}
    \begin{tabular}{c c c}
    \toprule
    Methods	& Measures	& \textit{p-value} \\
    \midrule
    Z3-25Eli, SMT-25Eli, IGA-25Eli, IAHP-25Eli	& \textit{Disagreement}	& \textit{p<0.05} \\
    Z3-50Eli, SMT-50Eli, IGA-50Eli, IAHP-50Eli	& \textit{Disagreement}	& \textit{p<0.05} \\
    Z3-100Eli, SMT-100Eli, IGA-100Eli, IAHP-100Eli	& \textit{Disagreement}	& \textit{p<0.05} \\
    \bottomrule
    \end{tabular}
    \end{scriptsize}
    \caption{ANOVA test comparing Z3 with SMT, IGA, and IAHP for different Elicited Pairs with 0\% Analyst Error.}
    \label{tab:RQ2:test}
\end{table}

\vspace{2mm}
\noindent \textbf{RQ3 (Robustness)} We conjecture that the Z3-based prioritization method is robust to errors made by the analyst. A simple stochastic model is used to simulate user rate at a varied percent of analyst errors, i.e., error$_{analyst}$ as 0\%, 5\%, 10\%, 20\%. For example, with 5\% error$_{analyst}$, responses from the analyst are 95\% of the times according to the GS and 5\% of the times reverse to the GS.

\begin{figure}[t!]
         \centering
         \includegraphics[trim={0.1cm 0.1cm 0.1cm 2cm},clip,width=1\textwidth]{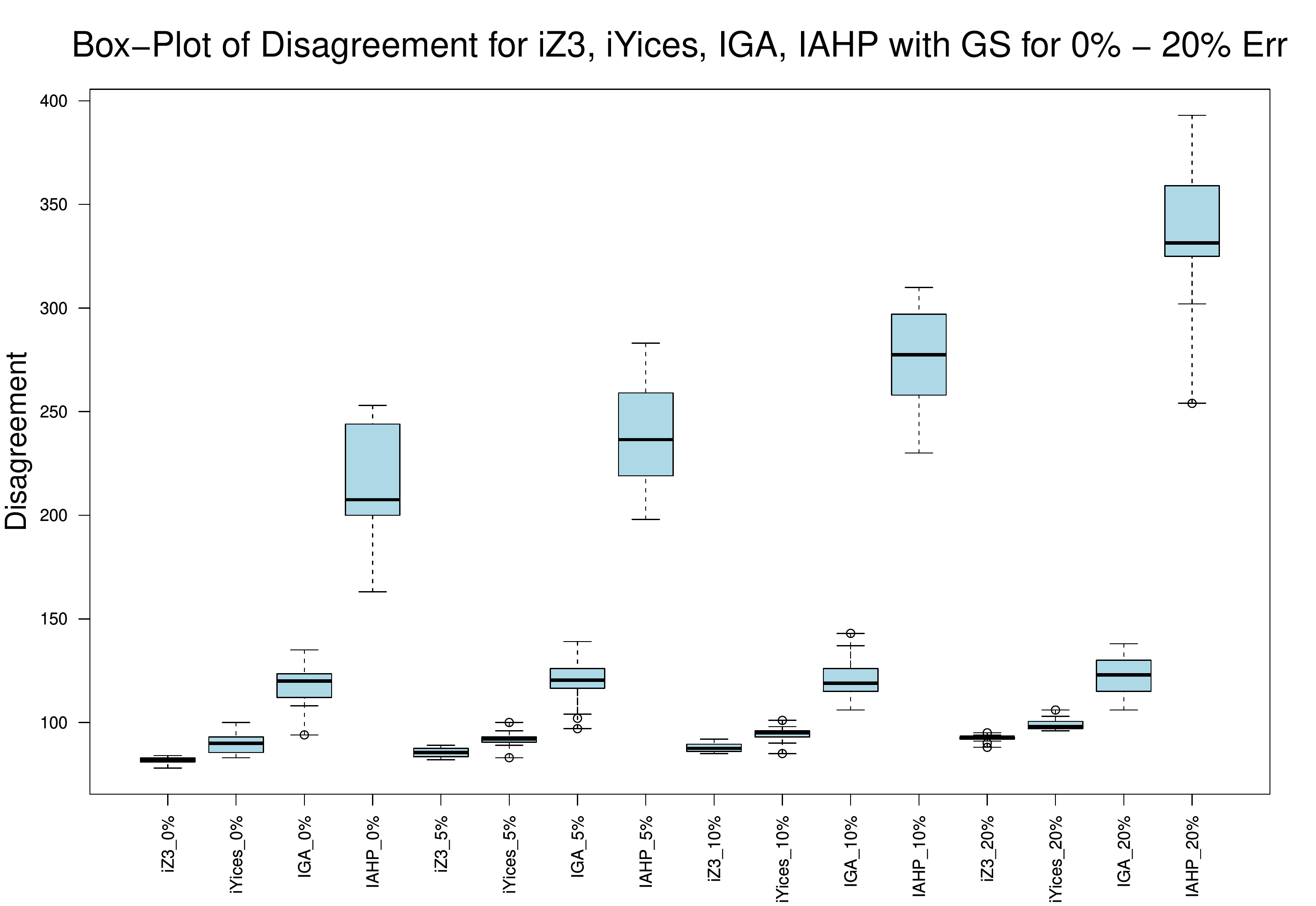}
        \caption{Robustness of Z3 compared to SMT, IGA, and IAHP  at different levels of Errors Made by the Analyst with 50 Pairs Elicited for ALL Scenario.}
        \label{fig:RQ3}
\end{figure}

Figure \ref{fig:RQ3} depicts the robustness of the Z3-based prioritization method compared to SMT (Yices), IGA, and IAHP at different levels of errors by the analyst with 50 elicited pairs for ALL scenario. A previous study showed that the SMT-based prioritization method is more robust than IGA and IAHP methods \cite{palma2011using}. Therefore, we only aim to show that the Z3-based solution is more robust than SMT, which would reflexively confirm that Z3 is more robust than IGA and IAHP methods. As Figure \ref{fig:RQ3} shows, the median disagreement using Z3 with 0\% error is 82; and the minimum disagreement using SMT with 0\% error is 92. Moreover, the minimum disagreements using Z3 with 5\%, 10\%, and 20\% error are 82, 85, and 88, respectively. These values using SMT for 5\%, 10\%, and 20\% error are 102, 102, and 101, respectively. Thus, the Z3-based prioritization method is more robust when it comes to errors made by the analyst. In particular, even after 20\% error, our Z3-based prioritization method outperforms the best rankings produced by error-free SMT (see Table \ref{tab:Results}).

\subsection{Discussion}

We positively answered three research questions based on the accumulated data from our experiments. As we showed for RQ1, our Z3-based interactive solution outperforms the non-interactive version. The more input from the analyst is considered (in terms of the number of elicited pairs), the better ranking of the requirements is produced. As reported for RQ2, our Z3-based prioritization outperforms its predecessor interactive requirements prioritization methods, e.g., SMT, IGA, and IAHP. The scale of improvement is statistically significant, as shown using ANOVA tests. We compared the performance of different methods in terms of disagreement and average distance against the gold standard. We also showed in RQ3 that our Z3-based interactive solution is more robust than the state-of-the-art interactive methods. The Z3-based solution outperforms its closest method, SMT, in terms of disagreement and average distance, although Z3 with 20\% error still produces a better ranking of the requirements than SMT with no error.

We also observed that both Z3 and SMT-based solutions perform a plateau after eliciting the maximum possible pairs by the analyst. However, when we elicit fewer pairs, i.e., 25 and 50, the Z3-based solution outperforms SMT. Thus, with less analyst input, Z3 outperforms SMT, which is also desired when there is a large number of requirements that may trigger a magnitude of comparisons to be made by the analyst. In other words, the Z3-based solution can minimize the effort and time required from the analyst. For instance, using the AHP, a well-known decision-making method, would require for ALL scenario with 49 requirements up to 1,176 comparisons made by the analyst. Nevertheless, when we compare our results with a maximum of 100 elicited pairs (in reality, we had a maximum of 84 pairs to be elicited given the available constraints), the performance of IAHP (a variant of AHP) is convincingly outperformed by all other interactive methods considered in this study. To conclude, our Z3-based interactive solution outperformed other state-of-the-art interactive methods regardless of the number of elicited pairs by the analyst.

\subsection{Threats to Validity}

The experiments are conducted on the ACube project, and thus, the findings may not be generalized to other systems. We considered four macro scenarios of the ACube project to generalize the findings and plan to conduct more experiments with other systems to minimize the threats to external validity. We used two metrics to answer our three research questions: disagreement and average distance. However, other metrics, e.g., hamming or levenshtein distance, could be used. Also, a more sophisticated error model in RQ3 could be used to better assess the robustness and significance of constraints, respectively. We performed statistical tests to confirm that the differences among the groups are significant.

\section{Conclusions and Future Work}\label{sec:Conclusions}
We introduced an interactive requirements prioritization method based on an efficient constraint solver Z3. Our method leverages pairwise requirements elicitation by an analyst. Various constraints, i.e., domain knowledge accumulated from requirements documents and analyst's knowledge, are encoded and serve as the input for our proposed method. The proposed method aims to reduce the effort (i.e., total number of pairwise comparisons) by requirements analyst while improving the accuracy of the final requirements ordering.

We conducted a range of experiments to validate our proposed prioritization method using a set of requirements from a real healthcare project ACube. We answered three research questions on the role of interaction, comparing with other state-of-the-art methods, and the robustness of the proposed method. The disagreement and average distance between the produced solution and the gold standard get lower when more pairs are elicited. Indeed, the Z3 constraint solver improved performance compared to SMT-based, IGA, and IAHP methods. Also, the Z3 constraint solver is more robust to analyst errors than other methods.

More investigation is required to find a more effective weighting scheme that would further improve the accuracy, i.e., minimize the disagreement against a total order of requirements. Also, we want to perform empirical studies with a human analyst to evaluate the usability and acceptability of the solutions produced by the proposed prioritization algorithm.



\bibliographystyle{splncs03}
\bibliography{CanadianAI2023/main}

\begin{thebibliography}{10}
\providecommand{\url}[1]{\texttt{#1}}
\providecommand{\urlprefix}{URL }

\bibitem{andrich2010acube}
Andrich, R., Botto, F., Gower, V., Leonardi, C., Mayora, O., Pigini, L.,
  Revolti, V., Sabatucci, L., Susi, A., Zancanaro, M.: {ACube: User-Centred and
  Goal-Oriented techniques}. Fondazione Bruno Kessler-IRST, Tech. Rep  (2010)

\bibitem{aurum2003fundamental}
Aurum, A., Wohlin, C.: {The fundamental nature of requirements engineering
  activities as a decision-making process}. Information and Software Technology
   45(14),  945--954 (2003)

\bibitem{avesani2005facing}
Avesani, P., Bazzanella, C., Perini, A., Susi, A.: {Facing scalability issues
  in requirements prioritization with machine learning techniques}. In: 13th
  IEEE International Conference on Requirements Engineering (RE'05). pp.
  297--305. IEEE (2005)

\bibitem{de2008z3}
De~Moura, L., Bj{\o}rner, N.: {Z3: An efficient SMT solver}. In: International
  conference on Tools and Algorithms for the Construction and Analysis of
  Systems. pp. 337--340. Springer (2008)

\bibitem{harker1987incomplete}
Harker, P.T.: {Incomplete pairwise comparisons in the analytic hierarchy
  process}. Mathematical modelling  9(11),  837--848 (1987)

\bibitem{in2002multi}
In, H.P., Olson, D., Rodgers, T.: {Multi-criteria preference analysis for
  systematic requirements negotiation}. In: Proceedings 26th Annual
  International Computer Software and Applications. pp. 887--892. IEEE (2002)

\bibitem{karlsson1996software}
Karlsson, J.: {Software requirements prioritizing}. In: Proceedings of the
  Second International Conference on Requirements Engineering. pp. 110--116.
  IEEE (1996)

\bibitem{karlsson1997cost}
Karlsson, J., Ryan, K.: {A cost-value approach for prioritizing requirements}.
  IEEE software  14(5),  67--74 (1997)

\bibitem{karlsson1998evaluation}
Karlsson, J., Wohlin, C., Regnell, B.: {An evaluation of methods for
  prioritizing software requirements}. Information and software technology
  39(14-15),  939--947 (1998)

\bibitem{khan2016repizer}
Khan, S.U.R., Lee, S.P., Dabbagh, M., Tahir, M., Khan, M., Arif, M.: {RePizer:
  a framework for prioritization of software requirements}. Frontiers of
  Information Technology \& Electronic Engineering  17(8),  750--765 (2016)

\bibitem{lauesen2002software}
Lauesen, S.: {Software requirements: styles and techniques}. Pearson Education
  (2002)

\bibitem{leffingwell2000managing}
Leffingwell, D., Widrig, D.: {Managing software requirements: a unified
  approach}. Addison-Wesley Professional (2000)

\bibitem{palma2011using}
Palma, F., Susi, A., Tonella, P.: {Using an SMT solver for interactive
  requirements prioritization}. In: Proceedings of the 19th ACM SIGSOFT
  symposium and the 13th European conference on Foundations of software
  engineering. pp. 48--58 (2011)

\bibitem{perini2012machine}
Perini, A., Susi, A., Avesani, P.: {A machine learning approach to software
  requirements prioritization}. IEEE Transactions on Software Engineering
  39(4),  445--461 (2012)

\bibitem{rodriguez2002models}
Saaty, R.W.: {The analytic hierarchy process—what it is and how it is used}.
  Mathematical modelling  9(3-5),  161--176 (1987)

\bibitem{sommerville1997re}
Sommerville, I., Sawyer, P.: {RE: A good practice guide}. John Wiley and Sons
  (1997)

\bibitem{5635176}
Tonella, P., Susi, A., Palma, F.: {Using Interactive GA for Requirements
  Prioritization}. In: 2nd International Symposium on Search Based Software
  Engineering. pp. 57--66 (2010)

\bibitem{tonella2013interactive}
Tonella, P., Susi, A., Palma, F.: {Interactive requirements prioritization
  using a genetic algorithm}. Information and software technology  55(1),
  173--187 (2013)

\bibitem{wiegers2013software}
Wiegers, K., Beatty, J.: {Software requirements}. Pearson Education (2013)

\end{thebibliography}

\end{document}